

\input phyzzx
\PHYSREV
\baselineskip 30pt
\sequentialequations
\pubnum{UTS-DFT-92-3}
\pubtype{HE}
\hfuzz 20pt
\mathsurround=2pt
\openup 2pt
\hfuzz 20pt
\def\h{H_{\mu\nu\rho}}
\def\H{H^{\mu\nu\rho}}
\def\b{B_{\mu\nu}}

\def\X{\dot X}
\def\Z{\dot Z}

\def\J{J_{\mu\nu}}
\def\J{J^{\mu\nu}}
\def\skr{S_{\rm KR}}
\def\gs{g_s}
\def\st{{1\over 2\pi\alpha'}}
\def\seq{S^{\rm eff.}}
\titlepage
\singlespace

\title{AXIONIC MEMBRANES }

\author{Antonio Aurilia\foot{E-Mail address: AAURILIA@CSUPOMONA.EDU}}
\address{Department of Physics\break
California State Polytechnic University\break Pomona, CA91768}

\andauthor{Euro Spallucci\foot{E-Mail
address:SPALLUCCI@TRIESTE.INFN.IT}}
\address{Dipartimento di Fisica Teorica\break Universit\`a di Trieste
\break INFN, Sezione di Trieste\break 34014-Trieste,Italy}

\submit{Physics Letters B}
\doublespace
\endpage
\abstract
A metal ring removed from a soap-water solution encloses a film of
soap which can be mathematically described as a minimal
surface having the ring as its only boundary.  This is known to everybody.

In this letter we suggest a relativistic extension of the above fluidodynamic
system where the soap film is replaced by a Kalb-Ramond gauge
potential $\b(x)$ and the ring by a closed string. The interaction between
the $\b$-field and the string current excites a new configuration of the
system
consisting of a relativistic membrane bounded by the string. We
call such a classical solution of the equation of motion an {\it axionic
membrane.} As a dynamical system, the axionic membrane admits a
Hamilton-Jacobi
formulation which is an extension of the H-J theory of
{\it electromagnetic strings}.
\vfill\eject
\REFS\Aa{%
Y.Nambu, Phys.Lett.{\bf 80B} (1979) 372;\hfill\break
J.L.Gervais and A.Neveau, Phys.Lett.{\bf 80B} (1979) 255;\hfill\break
A.M.Polyakov, Phys.Lett.{\bf 82B} (1979) 247; Nucl.Phys.{\bf B146} (1980)
171;
\hfill\break
Y.Nambu, Phys.Lett.{\bf 92B} (1980) 327.
}%
\REFSCON\Ab{%
H.B.Nielsen and P.Olesen, Nucl.Phys.{\bf B61} (1973) 45.
}%
\REFSCON\Ac{%
C.H.Tze and S.Nam, Ann.Phys.{\bf 193} (1989) 419; and refs.therein.
}%
\REFSCON\Am{%
See for instance, A.Aurilia and E.Spallucci `` The Role of Extended Objects
in Particle Theory and in Cosmology '' Proceedings of the Trieste Conference
on Super-Membranes and Physics in 2+1 Dimensions, Trieste 17-21 July
1989;
World Sci. Ed.
}%

\REFSCON\Af{%
Y.Nambu, Phys.Lett.{\bf 80B} (1979) 372;{\bf 92B} (1980) 327;
{\bf 102B} (1981) 149;
}%
\REFSCON\Ae{%
H.A.Kastrup, Phys.Lett.{\bf 82B} (1979) 237;\hfill\break
H.A.Kastrup and M.A.Rinke, Phys.Lett.{\bf 105B} (1981) 191;\hfill\break
M.Rinke, Comm.Math.Phys.{\bf 73} (1980) 265;\hfill\break
Y.Hosotani, Phys.Rev.Lett.{\bf 55} (1985) 1719.
}%
\REFSCON\Ap{%
F.J.Almgren, `` Plateau's Problem '' Benjamin, New York (1966)
}%
\REFSCON\Ag{%
M.Kalb and P.Ramond, Phys.Rev.{\bf D9} (1974) 2273.
}%
\REFSCON\Ah{%
P.A.M.Dirac, Phys.Rev. {\bf 74} (1948) 817
}%
\REFSCON\Ax{%
P.A.M.Dirac,Proc.R.Soc. London{\bf.A 268} (1962) 57
}%
\REFSCON\Ai{%
A.Aurilia and E.Spallucci, `` Gauge Field Representation and Hamilton-Jacobi
Theory of Relativistic Membranes`` Univ. of Trieste preprint, October 1991,
submitted to Nuclear Physics B.
}%
\REFSCON\Act{%
A.Chodos, C.B.Thorn, Nucl.Phys.{\bf B72} (1974) 509
}%
\REFSCON\Ar{%
A.Aurilia, Phys.Lett.{\bf B81}, (1979) 203
}%
\REFSCON\Aj{%
A.Aurilia and E.Spallucci `` String condensation and dynamical mass
generation
for $A_{\mu\nu\rho}$ '' in preparation.
}%
\refsend
\chapter{Introduction}

Relativistic quantum field theories admit both {\it local} (=particles) and
{\it non-local} (=extended objects) excitations. Remarkable examples of the
second kind are: string-like excitations in QCD\refmark{\Aa};
vortex-type solutions of the Higgs Model\refmark{\Ab};
membrane solutions of $\sigma$-models in $D=3,7,15$
spacetime dimensions\refmark{\Ac}.

On the other hand, one may take the point of view that relativistic
extended objects, regarded as submanifolds  embedded in spacetime,
are just as fundamental as the world-lines of point particles;
from this point of view, String-dynamics and Membrane dynamics are just
as fundamental as Electrodynamics. Thus, the distinction between
local quantum fields and extended structures in spacetime is not
as sharp as it would appear at first sight.

Motivated in part by our own investigations in particle physics and
cosmology
\refmark{\Am} where strings and domain walls are playing an increasingly
important role, and in part by the earlier work of Nambu
on the string dynamics of QCD\refmark{\Af},
{\it we wish to draw a precise correspondence
between the theory of extended objects governed by the Dirac-Nambu-Goto
action and the Hamilton-Jacobi theory of antisymmetric tensor gauge
fields.}

In order to make our discussion as sharp and focussed as possible, in
this paper we are primarily concerned with the Kalb-Ramond (KR)
formulation
of string dynamics. However, our discussion can be easily generalized to
manifolds of higher dimensionality.

The KR system involves a rank two gauge potential $\b(x)$ described by
the lagrangian
$I=-{1\over2\cdot3!}\h\H$, with $\h=\partial_{\,[\mu}B_{\nu\rho]}$.
Once coupled to the string
current, the $\b$ field mediates the interaction between the string
world-sheet surface elements, and the whole theory can be viewed as the
extension of Maxwell's electrodynamics in which the role of point-like
charges
is played by \underbar{closed strings}. The radiation field, in this case,
consists of massless spin-$0$ quanta. To the extent that such a field
possesses
a global Peccei-Quinn $U(1)$ symmetry, we shall refer to it as the
axion field.

Our specific purpose, then, is to show that the KR theory,
when suitably reformulated,
provides also a gauge invariant description of a membrane.bounded by a
closed string Such
extended structures, which we call {\it Axionic Membranes}, are
described by singular configurations of the $\h(x)$ field. They are the
counterpart of `` photonic '' or electromagnetic strings already known in
the literature\refmark{\Ae-\Af}. The underlying idea is simple: the world-
lines
of a pair of opposite point charges interacting with the electromagnetic
field define a world-sheet as the minimal surface bounded by them.
But the world-sheet can be alternatively
seen as the time evolution of an open string ending on the charges,
and the dynamics of this system can be reformulated in terms of string
coordinates.
In the same way, a closed string interacting with the Kalb-Ramond potential
can be viewed as the boundary of a two dimensional surface. From a more
physical point of view, the B-quanta exchanged between string elements
give
rise to an energy layer which, in the long wavelength limit,
can be approximated by a geometrical surface.
{\it Indeed,
we will show that, once written in terms of membrane variables, the
equations
of motion of the system are the same as the equations of motion for a
membrane having the original string as its only boundary.} This leads us to
the analogy with soap-films and minimal surfaces\refmark{\Ap}.

Everybody has sometime performed the simple experiment of dipping a
metal
ring into a soap-water solution. When removed from the solution the
ring encloses a film of soap. Such a thin layer is a minimal surface
having the ring as its only boundary. In our relativistic model {\it
the soap-film is replaced by the Kalb-Ramond field and the ring by a closed
string. The interaction between the local field and the one-dimensional
extended
object excites a new configuration of the system
consisting of a singular energy layer localized on an extremal
surface bounded by the string.}

\chapter{From String-dynamics to Membrane Theory}

We start from the Kalb-Ramond action coupled to a {\it closed string}
$$\eqalign{
S&\equiv S_{\rm KR}+S_{\rm INT.}+S_{\rm NG}\cr
\skr&=-{1\over 2\cdot3!}\int_{\cal M} d^4x\,\h(x)\H(x)\ ,\qquad
\h(x)=\partial_{[\mu}B_{\nu\rho]}(x)\cr
S_{\rm NG}&=-\st\int_K
d\tau d\sigma\sqrt{-{1\over2}\X^{\mu\nu}\X_{\mu\nu}}\equiv
\int_K d\tau d\sigma L_{\rm NG}\ .
\cr}
\eqno{(1)}
$$
$S_{\rm NG}$ is the string Nambu-Goto action, which is defined as follows:
we have a domain $K$ in the space of the parameters $s^a=(\tau,\sigma)$
which represent local coordinates on the lorentzian string manifold, and an
embedding $\Omega$
of $K$ in Minkowski spacetime ${\cal M}$, that is $\Omega:s\in
K\rightarrow
\Omega(s)\equiv X^\mu(s)\in {\cal M}$. We note, in particular, that the
tangent space $T(K)$ is mapped into the tangent space $T(\Omega)$ in ${\cal
M}$.
Thus, the tangent bi-vector in parameter space
$$
{\partial\over\partial\tau}\wedge{\partial\over\partial\sigma}
$$
is mapped into the tangent bi-vector\foot{ We shall also use the
shorthand notation
$$
{\partial(\dots)\over\partial\xi^0}\wedge{\partial X^\mu\over\partial\xi^1}
\dots\wedge{\partial X^\rho\over\partial\xi^d}\equiv
\{ (\dots), X^\mu,\dots X^\rho\}\ .
$$
}
$$
\X^{\mu\nu}={\partial X^\mu\over\partial\tau}\wedge
{\partial X^\nu\over\partial\sigma}\equiv
{\partial(X^\mu,X^\nu)\over\partial(\tau,\sigma)}
$$
at each point of the embedded submanifold $x^\mu=X^\mu(s)$
representing the string history ${\cal S}$ in spacetime.
We interpret the string world-sheet as the
only boundary of a 3-dimensional manifold ${\cal H}$, i.e.
${\cal S}=\partial{\cal H}$ .
Finally,
$$\eqalign{
S_{\rm INT}&=-{\gs\over2}\int_{\cal S}dX^\mu\wedge dX^\nu\b(X)\cr
& =-{\gs\over2}\int_{\cal M} d^4x
\int_Kd^2s\,\delta^{4)}\left(x-X(s)\right)\{X^\mu,X^\nu\}\b(x)\cr
&=-{\gs\over2}\int_{\cal M} d^4x\J(x)\b(x)\ ,\cr}
\eqno{(2)}
$$
which implicitly defines $\J(x)$ as the string current distribution
over the spacetime
manifold ${\cal M}$. Note that the coupling constant $\gs$ has units of
mass.

{\it Membrane variables} can be introduced into the theory by means of
Dirac's trick\refmark{\Ah}
which consists in defining the current as the divergence of a suitable
(singular) field:
$$\eqalign{
&\J\equiv\partial_\lambda J^{\lambda\mu\nu}\cr
&J^{\lambda\mu\nu}(x)\equiv\int_W d^3\xi\,\delta^{4)}\left(x-Z(\xi)
\right)\Z^{\lambda\mu\nu}\cr
&\Z^{\lambda\mu\nu}\equiv\{Z^\lambda,Z^\mu,Z^\nu\}=
{\partial(Z^\lambda,Z^\mu,Z^\nu)\over\partial(\tau,\sigma,\eta)}
\ .\cr}\eqno{(3)}
$$

Here $W$ is a domain in the space of the parameters
$\xi^a=(\tau,\sigma,\eta)$
having $K$ as boundary. Then, the world-tube ${\cal H}$ in spacetime
is the image of the map $Z:\,W\rightarrow {\cal M}$. In other words:
 $Z^\mu(\xi)\equiv Z^\mu(\tau,\sigma,\eta)$
represent the membrane
coordinates which parametrize the membrane
history ${\cal H}$ and are subject to the condition
$Z^\mu\Big\vert_{\partial\cal H}=X^\mu$.

Then, $S_{\rm INT}$ can be written as
$$
S_{\rm INT}={\gs\over 3!}\int_{\cal M} d^4x \h(x)J^{\mu\nu\rho}(x)\ .
\eqno(4)
$$
where $J^{\mu\nu\rho}$ represents the current distribution associated with
the 3-dimensional manifold of a membrane having the closed string
as its only boundary.

Note, however, that the action (1) does not contain a free membrane term.
Furthermore, after the above manipulations, the action depends on
$\b(x)$ {\it only} through the gauge invariant field strength $\h(x)$
and our objective is to show that $\h(x)$
provides a suitable {\it field representation of a geometric membrane}.
Variation of the action (1-4) leads to the following
set of field equations:
$$\eqalignno{
&\partial_\lambda H^{\lambda\mu\nu}=\gs\partial_\rho J^{\rho\mu\nu}
&(5a)\cr
&\{H_{\lambda\mu\nu}(Z),Z^\mu,Z^\nu\}=0 &(5b)\cr
&\{\Pi_{\mu\nu}(X),X^\nu\}={\gs\over2}\h(X)\X^{\nu\rho} &(5c)\cr
&\Pi_{\mu\nu}(X)\equiv \st{\X_{\mu\nu}\over\sqrt{-
{1\over2}\X_{\rho\sigma}
\X^{\rho\sigma}}}\ .&(5d)\cr }
$$
The Axionic Membrane is described by a {\it special} solution, say $\hat\H$,
of  eq.(5a): apart from the coupling constant, it is the current distribution
of the membrane,
$$
\hat H^{\lambda\mu\nu}(x)=\gs\int_W d^3\xi\,
\delta^{4)}\left(x-Z(\xi)\right)\{Z^\lambda,Z^\mu,Z^\nu\}
=\gs J^{\lambda\mu\nu}(x)\ .
\eqno{(6)}
$$
The solution (6) is singular and differs from zero only along the path
$x^\mu=Z^\mu(\xi)$ where the Bianchi identity $dH=0$ breaks down. In other
words the solution (6) represents a topological `` world-tube '' singularity
in Minkowsky space.


An {\it effective} theory for the membrane can be obtained by inserting
the solution (6) into the action (1), that is, by going on-shell
$$\eqalign{
S^{\rm eff.}&={\gs^2\over 2\cdot3!}\int d^4x\int d^3\xi'\,
\delta^{4)}\left(x-Z(\xi')\right)\Z^{'\,\mu\nu\rho}
\int d^3\xi\,\delta^{4)}\left(x-Z(\xi)\right)\Z_{\mu\nu\rho}+
S_{\rm NG}\cr
&={\gs^2\over 2\cdot3!}\int d^4x\delta^{4)}\left(x-Z(\xi)\right)
\int d^3\xi'\int d^3\xi\,\delta^{4)}\left(Z(\xi)-Z(\xi')\right)
\Z^{'\,\mu\nu\rho}\Z_{\mu\nu\rho}+
S_{\rm NG}\cr
&={\gs^2\over 2\cdot3!}
\int d^3\xi'\int d^3\xi\,\delta^{4)}\left(z-z'\right)
Z^{'\,\mu\nu\rho}Z_{\mu\nu\rho}+S_{\rm NG}\ ;\cr
z&=Z(\xi)\ ,\quad z'=Z(\xi')\ .\cr}
\eqno{(7)}
$$
At a point $P=Z^\mu(\tau,\sigma,\eta)$ 
consider a reference
frame such that the $z^1$-$z^2$ plane is tangent to the membrane at $P$,
and observe that a geometric surface with no thickness is only a
mathematical
idealization of a physical membrane. To be more realistic, one has
to consider a thin layer of
width $2a$ along the $z^3$ direction where the $\h(x)$ field is non
vanishing.
We take this physical requirement into account by considering the following
approximation of the $\delta$-function along the $z^3$ direction:
$$\eqalign{
\delta^{4)}\left(z-z')\right)&=\delta(Z^0-Z^{'0})\delta(Z^1-Z^{'1})
\delta(Z^2-Z^{'2})\delta(z^3-z^{'3})\cr
&={\delta^{3)}\left(\xi-\xi'\right)\over\sqrt{-
\gamma}}{1\over4a\sqrt\pi}\cr}
\eqno{(8)}
$$
where $\gamma=-{1\over3!}\Z_{\mu\nu\rho}\Z^{\mu\nu\rho}$ is the
determinant
of the induced metric on $\cal H$, and, by definition
$$
\delta(z^3-z^{'3})=\cases{{1\over4a\sqrt\pi}, & for $|z^3-z^{'3}|<2a$;\cr
0, &if $|z^3-z^{'3}|>2a$ .\cr}
\eqno{(9)}
$$
Now the $\xi'$ integration in eq.(7) can be easily done, and
one finds the effective action
$$
\seq=-\st\int_K d\tau d\sigma\sqrt{-
{1\over2}\X^{\mu\nu}(s)\X_{\mu\nu}(s)}
-\rho^{\rm eff.}
\int_W d\tau d\sigma d\eta\sqrt{-{1\over3!}\Z^{\mu\nu\rho}(\xi)
\Z_{\mu\nu\rho}(\xi)}\ .
\eqno{(10)}
$$
{\it which consists of the Nambu-Goto action for a string plus a Dirac term
\refmark {\Ax}for a membrane with an effective surface tension }
$$
\rho^{\rm eff}={\gs^2\over4a\sqrt\pi}\ .
\eqno{(11)}
$$

It may be useful to recall that in eq. (10) $s^a=(\tau,\sigma)$, $\xi^a=(\tau,
\sigma,\eta)$, so that, if $W=D\times R$ where $D$ is the unit disk
with $\eta= 1$ and $R$ is the time axis, then $K=S^1\times R$, and
$X^\mu(\tau,\sigma)=Z^\mu(\tau,\sigma,1)$. In this connection, it is
perhaps worthwhile to point out that eq.(10) represents the generalization
of the action for the string
with massive endpoints\refmark{\Act}. Accordingly we interpret $\seq$ as
describing a membrane with a ``~massive~'' stringlike boundary.

Our next step, now, is to show that the effective action (10) leads to the
remaining equations of motion (5b,c,d). First, we note that by taking into
account the smeared delta-function (8,9), the value of the field
strength on the membrane manifold is
$$
\hat\h(Z)={\gs\over 4a\sqrt\pi}{\Z_{\mu\nu\rho}\over\sqrt{-\gamma}}
\eqno{(12)}
$$
while along the string world-sheet the value of the H-field is the average
between its values inside and outside the membrane

$$
\hat\H(X)\equiv {\gs\over 8a\sqrt\pi}{\X^{\mu\nu\rho}\over\sqrt{-
\gamma}}\ .
\eqno(13)
$$
In both cases $H$ satisfies the Hamilton-Jacobi equation\refmark{\Ai}
$$
-{1\over3!}\hat\h\hat\H={\rm const.}\ .
\eqno(14)
$$
These are the expected values of the momentum conjugate to
$\Z^{\lambda\mu\nu}$, as we shall verify in the following.
According to the above relations, the system (5b, 5c, 5d) becomes
$$\eqalignno{
&\left\{{\Z_{\lambda\mu\nu}\over\sqrt{-\gamma}},Z^\mu,Z^\nu
\right\}=0
&(15)\cr
&\{\Pi_{\mu\nu}(X),X^{\nu}\}=
{\gs^2\over 16a\sqrt\pi}{\X_{\mu\nu\rho}\over\sqrt{-\gamma}}
\X^{\nu\rho}\ . &(16)\cr}
$$
These are the equations of motion for a membrane bounded by
a closed string interacting with the $\b$ field through a generalized
``Lorentz force'' term. Indeed, following ref.$(12)$, we apply
Hamilton's principle to the action $(10  )$ and find
$$\eqalign{
\delta\seq&=\int_K d\tau d\sigma \{\Pi_{\mu\nu},X^\nu\}\delta X^\mu-
\int_K d\tau d\sigma
{1\over2}\{\Pi_{\mu\nu\rho},X^\nu,X^\rho\}\delta X^\mu\cr
&+\int_W d\tau d\sigma d\eta \{\Pi_{\mu\nu\rho},Z^\nu,Z^\rho\}\delta
Z^\mu\cr}
\eqno{(17)}
$$
where
$$
\Pi_{\mu\nu\rho}(Z)={\partial L_{\rm NG}\over\partial\Z^{\mu\nu\rho}}=
-\rho^{\rm eff.}{\Z_{\mu\nu\rho}\over\sqrt{-
\gamma}}\quad\hbox{with}\quad
\Pi_{\mu\nu\rho}(X)=
-{\rho^{\rm eff.}\over 2}{\X_{\mu\nu\rho}\over\sqrt{-\gamma}}
\eqno{(18)}
$$
is the membrane {\it volume conjugate momentum}\refmark{\Ai}.
Comparing with eq.(14), and noting that
$$
-{1\over3!}\Pi_{\mu\nu\rho}\Pi^{\mu\nu\rho}=(\rho^{\rm eff.}){}^2
\eqno{(19)}
$$
we are led to identify the volume momentum of the membrane with the
singular axionic field:
$$
\hat\h(Z)=-{1\over\gs}\Pi_{\mu\nu\rho}(Z)\ .\eqno{(20)}
$$
Finally, since
$$\eqalignno{
&\{\Pi_{\mu\nu}(X),X^\nu\}={1\over2}\Pi_{\mu\nu\rho}(X)\X^{\nu\rho}
&(21a)\cr
&\left\{{\{Z^\lambda,Z^\mu,Z^\nu\}\over\sqrt{-\gamma}},Z^\mu,Z^\nu
\right\}=0 &(21b)\cr}
$$
the action $(10)$ reproduces the set of classical equations $(15-16)$.
Equations $(21)$ represent
the generalization of equations $(2.2)$, $(2.3a,3b)$ in ref.$(10)$ and
show that the string moves under the influence of a ``~generalized
Lorentz force~'' generated by the membrane, while eq.$(21.b)$ is the
same as the equation of motion for the body of a ``~massless~'' membrane.

Having achieved our intended objective, we wish to comment briefly on
the general pattern
which underlies the above discussion. Dirac's `` trick '',
eq.(3), which is central to our construction, in actual fact is no trick at
all.
Rather, it represents a special case of the geometric relation
$$
\partial_{\mu_1}J_{(p)}^{\mu_1\mu_2\dots\mu_p}(x)=
J_{(p-1)}^{\mu_2\dots\mu_p}(x)
\ . \eqno{(22)}
$$
Equation $(22)$ involves the set of five possible {\it p-chains}, or de Rham
current distributions, with support in Minkowski space,
$$
\eqalign{
&J_{(0)}(x)=\delta^{4)}\left(x-Z\right)\qquad
(p=0,\quad\hbox{by definition})\cr
&J_{(p)}^{\mu_1\dots\mu_p}(x)=\int_{\cal H}\delta^{4)}\left(x-Z\right)
dZ^{\mu_1}\wedge\dots\wedge dZ^{\mu_p}\cr
&\phantom{J_{(0)}^{\mu_1\dots\mu_p}(x)}\equiv
\int_{W^p}d^p\xi\,\delta^{4)}\left[x-Z(\xi)\right]
{\partial(Z^{\mu_1}\dots Z^{\mu_p})\over\partial(\xi^1\dots\xi^p)}\qquad
(p\ge 1)\ .\cr}
\eqno{(23)}
$$
Here ${\cal H}$ represents the world-track of the extended object
parametrized
by the map $\xi^a\rightarrow Z^\mu(\xi^a)$, $a=1,\dots p$.  According to
eq.$(22)$, the divergence
operation maps each p-chain into the $(p-1)$-chain associated with the
{\it boundary} of the world history of the extended object.
Thus, if there is no boundary, i.e. if the object is spatially
closed, then eq.$(22)$ yields zero.
Furthermore, a repeated application of the
divergence operation maps a p-chain to zero: the boundary of a boundary is
zero.

As mentioned in the introduction, the application of eq.(22) for $p=2$ allows
the formulation of electrodynamics in terms of world-sheets and this leads
to the concept of electromagnetic or `` photonic '' strings.

The KR system represents the next level in the linkage among geometric
structures: a closed string may constitute the boundary of a membrane;
eq.(3) is now seen as a special case of eq.(22) for $p=3$ and eqs.(3-5)
imply that on the world-tube of the membrane $\H\sim
J_{(3)}^{\mu\nu\rho}$
( axionic membranes ).

The predictive power of this geometric scheme is seen in the next level
where
a closed membrane is coupled through the 3-chain $J_{(3)}^{\mu\nu\rho}$ to
a potential $A_{\mu\nu\rho}$ whose field strength is the 4-form
$F_{\mu\nu\rho\sigma}=\partial_{\,[\mu}A_{\nu\rho\sigma]}$. The closed
membrane
constitutes the boundary of a `` bag '' described by a world-hypervolume
to which we associate the 4-chain $J_{(4)}^{\lambda\mu\nu\rho}$. Then,
within the world-hypervolume,
$F^{\lambda\mu\nu\rho}=g_MJ_{(4)}^{\lambda\mu\nu\rho}$, where $g_M$
represents
the coupling constant of the membrane.

In terms of the embedding coordinates, one would expect
$$
F^{\lambda\mu\nu\rho}=g_M{\{Z^\lambda,Z^\mu,Z^\nu,Z^\rho\}\over
\sqrt{-{1\over4!}\Z^{\lambda\mu\nu\rho}\Z_{\lambda\mu\nu\rho}}}
$$
so that $F^{\lambda\mu\nu\rho}$ satisfies the Hamilton-Jacobi equation
$$
-{1\over 4!}F^{\lambda\mu\nu\rho}(Z)F_{\lambda\mu\nu\rho}(Z)=g^2_M\ .
$$
Note that $F^{\lambda\mu\nu\rho}$ represents nothing more than a constant
volume
term. This interpretation is consistent with our previous formulation of the
`` bag-model '' of hadrons. In flat spacetime the effect of
$F^{\lambda\mu\nu\rho}$-field is to induce a ``~bag constant~'' into the
field equations\refmark{\Ar}; in curved spacetime the effect of the
$F^{\lambda\mu\nu\rho}$-field is to introduce a cosmological constant into
Einstein's equations\refmark{\Am}.

We conclude by noting that a remarkable interplay between strings and
membranes
also appears at the quantum level. In the framework of string
field theory, membranes appear as topological singularities coupled
to the Kalb-Ramond field strength in a non trivial vacuum
state\refmark{\Aj}.
There, the singular field
$J^{\mu\nu\rho}$ arises  as a `` vorticity source '' for the vector phase
$\theta_\mu(x)$ of the string functional, that is
$J_{\mu\nu\rho}\prop\partial_{\,[\mu}\partial_\nu\theta_{\rho]}$. The
membrane
represents the spacetime sub-manifold where the Bianchi identities for
$\theta_\mu(x)$ are violated. This is consistent with our identifications (6)
and (20).

\refout
\bye